\documentclass[a4paper,prb,onecolumn,amsmath,amssymb]{revtex4}
\usepackage[latin9]{inputenc}
\usepackage[T1]{fontenc}
\usepackage{graphicx}
\usepackage{enumerate}
\usepackage{amsmath}
\usepackage{esint}
\usepackage{fullpage}
\usepackage{color}
\usepackage{epsfig}
\usepackage{epstopdf}

\begin{document}
	
\title{   On the Exciton Fine-Structure of Transition-Metal Dichalcogenides Mono-Layers}

\author{Pierre Gilliot, Mathieu Gallart, and Bernd Hönerlage}
\affiliation{\\~~~~~~~\\IPCMS,  UMR 7504, CNRS and Universit\'e de Strasbourg,\\ 23, rue du L{\oe}ss, F-67034 Strasbourg, France }

\maketitle
\section*{    Abstract}

In order to discuss the exciton fine-structure of transition-metal dichalcogenides mono-layers, excitons are first defined in the subspace of electron- and hole states, including the lowest conduction band (LCB) and the uppermost valence band (UVB). Both bands are spin degenerate at the $\Gamma$-point. All other states are neglected. The resulting exciton states are analyzed in the framework of an invariant expansion of a model Hamiltonian: The spin-orbit coupling in the conduction- and valence band is simulated by introducing a fictive magnetic field, giving rise to a splitting of the electron- and hole states outside the $\Gamma$-point. Then the electron-hole exchange-interaction is introduced into the exciton Hamiltonian. It is due to the fact that electron and hole are indistinguishable particles in the exciton problem. In $D_{3h}$ crystal symmetry this electron-hole exchange-interaction has two  different contributions: While a first term accounts for an energy re-normalization of all exciton states, a second term does not influence the optical active (spin-singlet) states but affects only the optical inactive (spin-triplet) states, which become mixed in-between the different exciton series. \\

%\newpage

%\begin{center}
\section{ Introduction   }
%\end{center}

In this paper, we intend to discuss the exciton fine-structure of transition-metal dichalcogenides (TMD) mono-layers such as $MoSe_2$. While bulk $MoSe_2$ is an indirect-gap semiconductor crystal, $MoSe_2$ mono-layers possess a direct energy gap at the $K_{\pm}$-points of the two-dimensional Brillouin zone (see Ref. \cite{DXiao_2012}). In a first approximation one can construct exciton states from the lowest conduction band (LCB) and the uppermost valence band (UVB) of the semiconductor crystal and neglect all other states. An exciton is formed if an electron is excited from the valence band to the conduction band, leaving behind an unoccupied place in the otherwise completely filled valence band. This "missing electron" state is transformed into a "hole" state in the exciton problem. Electron and hole interact with each other via Coulomb interaction and form bound states. These exciton states are either optically active (bright) or inactive (dark) depending on their coupling to the electromagnetic radiation field. Because of the strong spin-orbit coupling acting in the valence band of TMD crystals two exciton series (called "A" and "B" series according to their energy) are observed in optical measurements on these materials. Their fine-structure is determined by the details of the electron-band structure and the symmetry properties of the layers. 

Bulk $MoSe_2$ crystals are built from ionic covalently bound $Se-Mo-Se$ units, which form hexagonal planes, referred to as "mono-layers" in the following. On the contrary to the intra-layer binding, adjacent mono-layers are only weekly coupled with each other through Van der Waals forces. Due to the different coupling mechanisms inside the same and in-between different layers, bulk $MoSe_2$ is a laminar, layered semiconductor, having $D_{6h}$ point-group symmetry (see Ref. \cite{ZYZhu_2011}).
  TMD semiconductors possess an indirect energy band gap, showing only rather weak photoluminescence (PL) emission. The energy of the indirect gap of $MoSe_2$ is about 1.29 eV at room temperature. (See Ref. \cite{KFMak_2010})

When decreasing the film thickness to ultra-thin bulk crystals, the indirect-gap energy increases successively as a function of the number of mono-layers because of spatial confinement. It reaches values of over 1.90 eV under ambient conditions in the case of mono-layers. (See Ref. \cite{KFMak_2010})
This increase of the indirect-gap energy is much larger than that of the direct-gap energy, which increases only by 0.1 eV to about 1.8 eV.
Thus, the electron band structure of $MoSe_2$ shows a crossover from an indirect- to a direct-gap semiconductor material in the limit of  mono-layer samples. (For a recent review on excitons in atomically thin TMDs see e.g. Ref.  \cite{GWang_2017})

 Comparing mono-layers to bulk material one remarks that the prismatic coordination inside the mono-layer is maintained, while the inversion symmetry operation in-between adjacent layers does no longer exist. Thus the point-group symmetry of the layer is reduced from $D_{6h}$ to  $D_{3h}$ symmetry. (See Ref. \cite{ZYZhu_2011})
It is important to notice that the irreducible representations of the  $D_{3h}$ point double group are either non degenerate or two times degenerate, which gives rise to spin-degenerate electron states at the $\Gamma$-point, even when including spin-orbit coupling in band-structure calculations. (See Ref. \cite{ZYZhu_2011}) 

The direct energy gap of transition-metal dichalcogenides mono-layers is situated at the $K_{\pm}$-points of the two-dimensional hexagonal Brillouin zone, which are situated at its edges. $K_{+}$ and $K_{-}$ points are connected to each other by time reversal symmetry. The finite wave vector breaks, however, the full point-group symmetry of the $\Gamma$-point. Then, the group-symmetry of the wave-vector at the band edges (at the $K_{\pm}$-points) is $C_{3h}$, i.e. their symmetry is lower than that at the $\Gamma$-point since three twofold $C_{2}$ rotations that lie in the horizontal mono-layer plane and the mirror reflection planes containing these $C_{2}$ rotations are missing \cite{Koster_1960, MMGlazov_2014, YSong_2013}. As a consequence and on the contrary to the situation at the $\Gamma$-point, all irreducible representations of the $C_{3h}$ point double group are non degenerate. (See Ref. \cite{ZYZhu_2011})
 The point-group symmetry being reduced at the $K_{\pm}$-points, spin-orbit interaction  leads to an important splitting of states that are degenerate at the center of the Brillouin zone. 

In the semiconductor ground state all electron valence-band states are filled and all conduction-band states are empty. When adding some energy to a valence-band electron, it can be excited from the valence- to the conduction band, leaving behind an unoccupied state in the valence band. This situation corresponds to the excitation of an electron-hole pair called "exciton", which is an electronic elementary excitation. It can be looked upon as a quasi-particle, i. e. it may be characterized by an energy, a wave-vector, an angular momentum, etc. In addition, excitons have a fine structure that depends on the multiplicity of the electronic states and on the interactions to which they are subjected.

When photo-excited, direct-gap materials show usually an important PL emission intensity since electron-hole pairs can directly recombine with each other without involving further scattering processes with phonons or crystal imperfections. Because of their direct energy gap, transition-metal dichalcogenides mono-layers have a much higher luminescence quantum yield than the (indirect gap) bulk material. In addition, optical transitions are mainly observed in the vicinity of critical points, where the electron density of states is high. Often, as it is the case in TMD mono-layers, they are situated at high symmetry points of the crystal structure. Since exciton states are constructed from conduction- and valence-band electron states the emitted PL contains important information about the electron- and exciton-fine structure, which we will discuss in the following. (See Ref. \cite{KFMak_2010, HDery_2015})

Different methods (as  $\textbf{k}$ $\cdot$ $\textbf{p}$ perturbation theory \cite{MMGlazov_2014, DXiao_2012, HDery_2015, TYu_2014}, direct model calculations \cite{HYu_2014}, ab initio Bethe-Salpeter equation method  \cite{DYQiu_2015, JPEche_2016} etc.) have been used in literature to discuss the exciton fine-structure in TMD mono-layers. These calculations concentrate mainly on the  A-exciton ground state since it is often energetically well separated from the B-exciton states. (For recent review articles see e. g. Ref.  \cite{MMGlazov_2014, GWang_2017} and references cited therein.) Very interesting is also an approach where "bright" and "dark" exciton states (for this notation see chapter III) are constructed from electron- and hole states at the  $K_{\pm}$ critical points \cite{CRobe_2017}. The structure of the A-exciton ground state is then derived from the symmetry properties of the exciton states for important spin-obit splittings of valence-band electron-states. The exciton fine-structure is shown to be related to the direct- and exchange Coulomb interaction between electron and hole and to the spin-orbit splitting of the conduction band.

We apply in this paper a similar method by constructing a model exciton Hamiltonian, to which an invariant expansion is applied. On the contrary to Ref.\cite{CRobe_2017} we define, however, exciton states in the product space of electron- and hole subspace, in which we carry out the invariant expansion of the model Hamiltonian. After solving the corresponding Schrödinger equation, exciton eigenvalues and eigenfunctions are obtained. This method enables us to analyze in detail the mixing structure of the electron-hole pair states (including the spin- and and valley configuration as well as the A- and B-exciton states), the importance of the different Coulomb exchange interaction terms, and to establish the role that the electron spin-orbit coupling plays.

	\section{  Determination of an Effective Exciton Hamiltonian in $MoSe_2$ Mono-Layers  }
\setcounter{equation}{0}

  In semiconductors the exciton fine structure is governed by the symmetry properties of the crystal and of the atomic wave functions, which are built into the electronic Bloch functions. To analyze the exciton properties we construct a model Hamiltonian that respects the spatial and temporal symmetry properties of the crystal. The Hamiltonian is then developed into terms which remain invariant under the symmetry operations of the crystal and under time reversal. This method has been described in detail in Ref. \cite{Honer_invar_2018}.
  
  As mentioned above, in addition to the crystal symmetry, one has also to consider the temporal transformation properties of operators, which show up in the model Hamiltonian. We denote by  $\hat{K}^+$ ($\hat{K}^-$) the symmetry properties of operators that remain invariant (change their sign) under time reversal. (In order to avoid confusion with the high-symmetry points $K$ of the Brillouin zone, $\hat{K}$ stands here for the operation of Kramers' conjugation, i. e. for time reversal.) 
  
  Let us now determine an effective Hamiltonian describing the exciton-fine structure in transition-metal dichalcogenides  mono-layer semiconductors such as $MoSe_2$. In this material both, conduction- (subscript "e") and valence-band states  (subscript "v") are made up from a hybridization of atomic p- and d orbitals, originating from the Se- and Mo atoms, respectively. The energy bands are spin degenerate by symmetry at the  $\Gamma$-point but may split (due to the symmetry-breaking properties of a finite wave-vector) through spin-orbit interaction.  (See Ref. \cite{ZYZhu_2011}) 
 
We first consider the two-fold degenerate conduction-band states at the $\Gamma$-point. All other electron states are neglected. In order to describe this two-dimensional subspace we introduce an effective "pseudo-spin operator" $\boldsymbol{\sigma_e}$ with $\sigma_e = 1/2$, which operates only on the conduction-band states. It is given by the Pauli-spin matrices $\sigma_e^i$ with $i = (x, y, z)$, which are chosen to span the subspace of the considered conduction-band electron-states and are used to construct the model Hamiltonian:
 
 \begin{equation}  
\sigma_e^x  =   \begin{pmatrix} 0 & 1 \\ 1 & 0 \end{pmatrix} ;              \sigma_e^y  =   \begin{pmatrix} 0 & -i \\ i & 0 \end{pmatrix} ;       \sigma_e^z  =   \begin{pmatrix} 1 & 0 \\ 0 & -1 \end{pmatrix}  
\label{eqII1}
\end{equation}

In the invariant development of the Hamiltonian we use operators that are adapted to the crystal symmetry, i. e. the operators transform like irreducible representations of the point double group of the considered crystal structure. According to Ref. \cite{Koster_1960} $\sigma_e^z$ transforms as the irreducible representation $\Gamma_2$ in systems with $D_{3h}$ point-group symmetry. Instead of the Pauli matrices
$ \sigma_e^x $ and $ \sigma_e^y$ we further choose their linear combinations

\begin{equation}
\begin{split}
\sigma_e^{61} = (\sigma_e^x - i \sigma_e^y )/ 2 =  \begin{pmatrix} 0 & 0 \\ 1 & 0 \end{pmatrix}   \\
\text{and }\\
\sigma_e^{62} =  - ( \sigma_e^x + i \sigma_e^y ) /2 =  \begin{pmatrix} 0 & -1 \\ 0 & 0 \end{pmatrix} ,\\
\end{split}
\label{eqII2}
\end{equation}
as basis matrices, since they transform as the elements of the two-dimensional irreducible representation  $\Gamma_6$.

The Pauli-spin matrices in equ. (II.1) transform like the components of an angular-momentum operator, which involves first-order time-derivatives. Therefore the Pauli matrices have $\hat{K}^-$ transformation symmetry under time reversal. 

In addition to $\sigma_e^z$ and the linear combinations of Pauli matrices in equ. (II.2) we introduce the unit matrix $1_e$ as a fourth basis matrix. The unit matrix $1_e$  is given by:

\begin{equation}
1_e =   \begin{pmatrix} 1 & 0 \\ 0 & 1 \end{pmatrix}    
= (1/3)[ ( \sigma_e^x )^2 + ( \sigma_e^y )^2 + (\sigma_e^z)^2 ] = (1/3) (\boldsymbol{\sigma_e} )^2  .
\end{equation}
Since the unit matrix $1_e$ does not depend on time (applying twice time reversal to an operator [here $\sigma_e^i$, i = (x, y, z)] restores its initial time dependence) this operator is invariant under time reversal and its overall symmetry is therefore noted ($\Gamma_1, \hat{K}^+$). 

 The matrices ($\sigma_e^{z}, \sigma_e^{61}, \sigma_e^{62}$) and the unit matrix $1_e$ are linear independent matrices. They span together a subspace in which the Hamiltonian describing the conduction-band electron is defined. As discussed in  Ref.  \cite{Honer_invar_2018, Bir_Pikus_1971} time independent model Hamiltonians have to be invariant under all symmetry operations of the crystal point group and with respect to time reversal, i. e. they have to transform as ($\Gamma_1, \hat{K}^+$). Then, in the absence of symmetry-breaking interactions, the Hamiltonian $H^{e}$ acting on the conduction-band electron in the two-dimensional subspace has the form: 
 \begin{equation}
H^{e} = E^{e} 1_{e} 
\end{equation}

where the spin-up and spin-down states ($\alpha_e$, $\beta_e$) are the eigenstates  of the Hamiltonian and ($E^e_{\alpha}$, $ E^e_{\beta}$) are its eigenvalues, obeying to:
 
 \begin{equation}
 E^e_{\alpha} = E^e_{\beta}
 \end{equation}
The pseudo-spin functions ($\alpha_e$, $\beta_e$) represent the wave functions of the electron in the conduction band, which are constructed from the atomic wave functions. They posses an orbital part, indicate the spin state ($\alpha, \beta$ for spin-up and spin-down), and have a total symmetry that is compatible with the point group of the crystal. $H^{e}$ describes all interactions that leave the states unchanged under all symmetry operations. No other term may appear in a Hamiltonian because of this symmetry condition.

Let us now consider the spin-orbit coupling in our model Hamiltonian describing the conduction-band electron-states. Because of the crystal symmetry the states are degenerate at the $\Gamma$-point. At finite wave-vectors $\textbf{Q}_e$, however, the full point-group symmetry is broken. Then, spin-orbit interaction gives rise to a splitting of the states, which depends on the direction and on the absolute value of the wave-vector. (See Ref. \cite{ZYZhu_2011})

The spin-orbit splitting can now be simulated through a fictive magnetic field $\textbf{B}^e(\textbf{Q}_e)$ in the following way: In crystals with $D_{3h}$ point-group symmetry a magnetic field $\textbf{B}^e = [0, 0, B_z^e]$ along the z-direction transforms (as the Pauli matrix $\sigma_e^z$ does) according to ($\Gamma_2, \hat{K}^-$).  (See Ref. \cite{Koster_1960})
Then, the product:
 \[
g^e \mu_B \textbf{B}^e  \sigma_e^z 
 \]
($g^e$ denoting the Landé factor of the conduction-band electron and $\mu_B$ the magneton of Bohr) transforms according to ($\Gamma_1, \hat{K}^+$), i. e. it has the transformation properties required for a Hamiltonian. (See Ref. \cite{Koster_1960})
Thus, in general, a magnetic field gives rise to a splitting between the two conduction-band states. The fictive magnetic field $\textbf{B}^e(\textbf{Q}_e)$ introduced here is now adjusted 
to a value that simulates the spin-orbit splitting function $\Delta_{so}^e (\textbf{Q}_e)$ of the conduction-band states at the considered wave-vector $\textbf{Q}_e$. $\Delta_{so}^e(\textbf{Q}_e)$ has been calculated e. g. in Ref. \cite{ZYZhu_2011}
for the conduction band of $WSe_2$ in the ($\Gamma$ - K) direction. Introducing
\[
  \Delta_{so}^e(\textbf{Q}_e) = 2 g^e \mu_B \textbf{B}^e(\textbf{Q}_e) = 2 a_{so}^{e}(\textbf{Q}_e) 
\]
the spin-orbit coupling is accounted for in our model Hamiltonian through the term: 

\begin{equation}
H^{e}_{so}(\textbf{Q}_e) = a_{so}^{e}(\textbf{Q}_e)\sigma_e^z =  \begin{pmatrix}  a_{so}^{e}(\textbf{Q}_e) & 0 \\ 0 & -  a_{so}^{e}(\textbf{Q}_e) \end{pmatrix} . 
\end{equation}

Our fictive magnetic field  $\textbf{B}^e(\textbf{Q}_e)$ simulates entirely the action of the spin-orbit coupling on the conduction-band electron-states. Attention has to be payed, however, that this fictive field is not constant (i. e. it does not behave like an external field) but it varies in function of the wave-vector $\textbf{Q}_e$. $a_{so}^{e}(\textbf{Q}_e)$ takes its maximum value at the K-points of the Brillouin zone where the direct energy gap is situated and  $a_{so}^{e}(\Gamma) \equiv 0 $ at the  $\Gamma$-point. The spin-orbit coupling and therefore  $a_{so}^{e}(\textbf{Q}_e)$ obey to the symmetry relation:
\[
 a_{so}^{e}(-\textbf{Q}_e) = -  a_{so}^{e}(\textbf{Q}_e) 
\]

\begin{figure}
	\begin{center}
		\includegraphics[width=15cm]{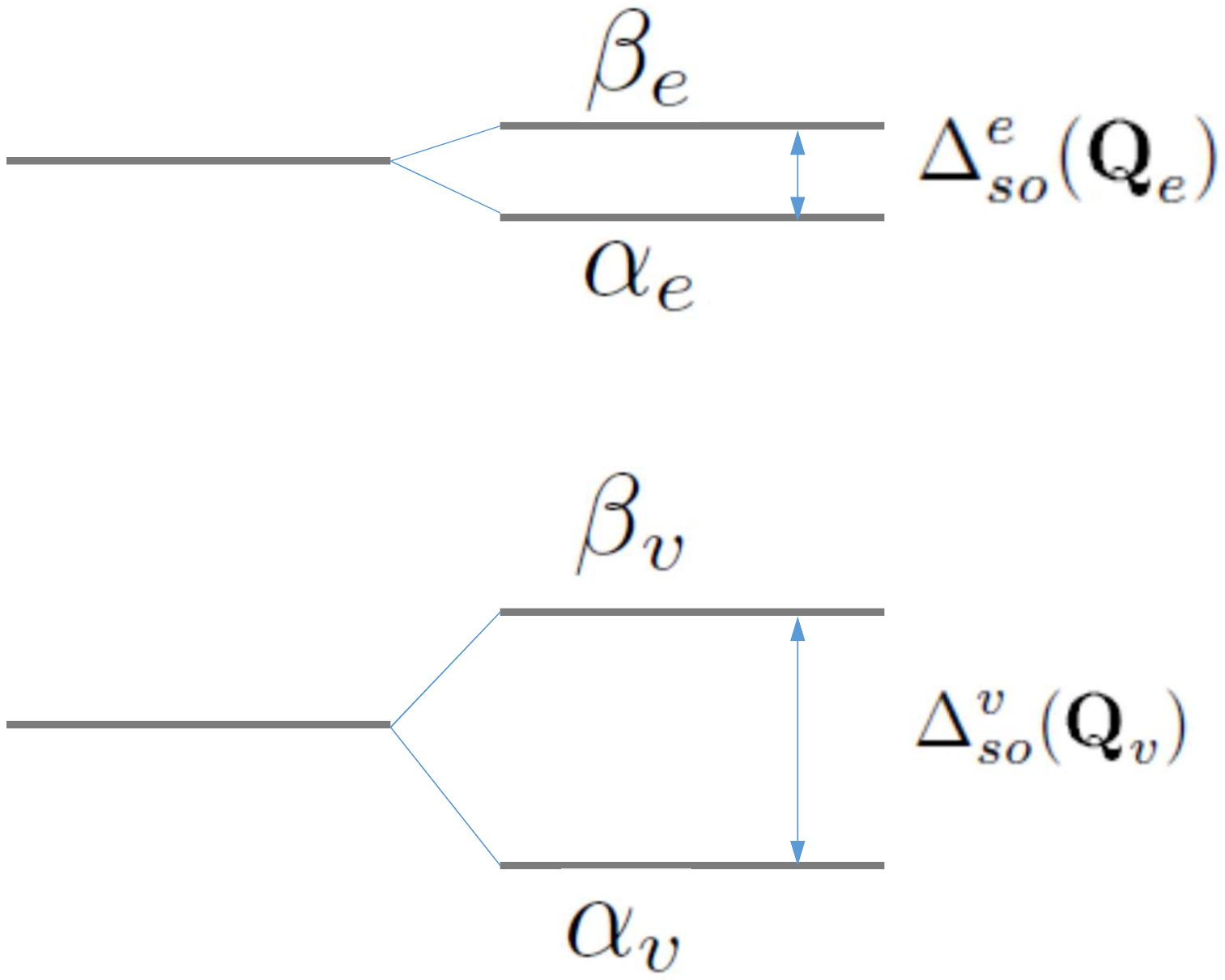}
	\end{center}
	\caption{Conduction and valence band structure resulting from the spin-orbit coupling.
	}
\end{figure} 
 
 The same construction as for the conduction-band is now applied to the valence-band states (subscript "v"). Their wave functions originate from the same type of atomic orbitals and these states are also two-times degenerate at the $\Gamma$-point. Similarly we introduce Pauli matrices

 \begin{equation}  
\sigma_v^x  =   \begin{pmatrix} 0 & 1 \\ 1 & 0 \end{pmatrix} ;              \sigma_v^y  =   \begin{pmatrix} 0 & -i \\ i & 0 \end{pmatrix} ;       \sigma_v^z  =   \begin{pmatrix} 1 & 0 \\ 0 & -1 \end{pmatrix}  
\end{equation}
which act only on the valence-band states and give then rise to the basis matrices that span the valence-band subspace:

\begin{equation}
\begin{split}
\sigma_v^z  =   \begin{pmatrix} 1 & 0 \\ 0 & -1 \end{pmatrix} ,\\
\sigma_v^{61} = (\sigma_v^x - i \sigma_v^y )/ 2 =  \begin{pmatrix} 0 & 0 \\ 1 & 0 \end{pmatrix} \text{and } 
\sigma_v^{62} =  - ( \sigma_v^x + i \sigma_v^y ) /2 =  \begin{pmatrix} 0 & -1 \\ 0 & 0 \end{pmatrix},  \\
\text{and } \\
1_v =   \begin{pmatrix} 1 & 0 \\ 0 & 1 \end{pmatrix}    
= (1/3)[ ( \sigma_v^x )^2 + ( \sigma_v^y )^2 + (\sigma_v^z)^2 ] = (1/3) (\boldsymbol{\sigma_v} )^2  
\end{split}
\label{eqII8}
\end{equation}
 where the pseudo-spin states ($\alpha_v$, $\beta_v$) are the eigenfunctions and ($E^v_{\alpha}$, $ E^v_{\beta}$) the eigenvalues of the corresponding effective Hamiltonian $H^{v}$ :
  \begin{equation}
 H^{v} = E^{v} 1_{v} .
 \end{equation}
($\alpha_v$, $\beta_v$) are again pseudo-spin functions which represent the valence-band wave functions (orbital part and spin) of the missing electron whose detailed form and symmetry has to be specified. The valence-band states are also degenerate in energy at the  $\Gamma$-point, i. e.
 
  \[
 E^v_{\alpha} = E^v_{\beta} .
 \]
 
  One may consider spin-orbit coupling inside the valence band in the same way as discussed above through another fictive magnetic field $\textbf{B}^v(\textbf{Q}_v)$, giving rise to a splitting of 2$a_{so}^{v}(\textbf{Q}_v)$ in-between the valence-band states. Then the effective spin-orbit Hamiltonian of the valence-band subspace $H^{v}_{so}$ reads: 
 
 \begin{equation}
 H^{v}_{so}(\textbf{Q}_v) =a_{so}^{v}(\textbf{Q}_v)\sigma_v^z =  \begin{pmatrix} a_{so}^{v}(\textbf{Q}_v) & 0 \\ 0 & - a_{so}^{v}(\textbf{Q}_v) \end{pmatrix}  
 \end{equation} 
 and
 \[
 \Delta_{so}^v(\textbf{Q}_v) = 2 a_{so}^{v}(\textbf{Q}_v) 
 \]
 denotes the spin-orbit splitting function $\Delta_{so}^v(\textbf{Q}_v)$ of the valence-band states.
  
 The fictive magnetic fields $\textbf{B}^e(\textbf{Q}_e)$ and $\textbf{B}^v(\textbf{Q}_v)$ (acting in the conduction- and valence band sub-spaces, respectively) lead to independent splittings of the bands when varying the wave-vector, but evidently they vanish at the $\Gamma$-point where both bands are degenerated by symmetry.

 Excitons are defined in the product space of conduction- and valence-band states that we will discuss in the following. Therefore the Kronecker product of conduction- and valence-band spin matrices determines a basis of the fourfold degenerate exciton-ground state in the matrix representation of its Hamiltonian. These exciton states are determined from the product of conduction- and valence-band electron-wave functions, which are given here in the basis:
 
  \begin{equation}
\arrowvert\beta_e\beta_v\rangle,\arrowvert\beta_e\alpha_v\rangle,  
\arrowvert\alpha_e\beta_v\rangle, \arrowvert\alpha_e\alpha_v\rangle \label{eqII11}
\end{equation} 
 
As discussed in detail in Ref. \cite{Honer_invar_2018}
the basis functions given above can be easily transformed into a basis of exciton states by replacing the valence-band states by hole states (denoted by the subscript "h"), i. e. by the Kramers' conjugated form of the valence-band wave-functions. (This transformation results in replacing the orbital part of the valence-band wave-function $w_v$ by its complex conjugated expression, i. e. $w_h = w_v^*$. Concerning spins, $\alpha_v$ has to be replaced by $\beta_h$ and $\beta_v$ by -$\alpha_h$. When considering optical transitions it is easier, however, to discuss the excited states in terms of conduction- and valence-band electron-states and to introduce excitons formed from electrons and holes only at the end after the  transition selection rules have been established.) 

 Considering only the direct exciton-binding energy $a_{d}$ due to the Coulomb interaction between electron and hole but neglecting spin-orbit coupling, electron-hole exchange interactions, and all symmetry breaking effects for the moment, the effective direct exciton Hamiltonian $H^{ex}_d$ is described by the four-dimensional unit matrix in this pseudo-spin product space: 
\begin{equation}
H^{ex}_d = a_{d} 1_{e} \otimes  1_v 
\label{eqII12}
\end{equation} 

  Let us now include in our Hamiltonian spin-orbit interaction \cite{AKorm_2013} in addition to the direct exciton-binding energy (but we still neglect all other symmetry breaking effects and electron-hole exchange). First we remember that exciton states are constructed from packets of conduction- and valence-band electron-wave functions, which are extended in momentum space. In our model of spin-orbit interaction the quasi-particle wave-vectors determine the fictive magnetic fields $\textbf{B}^e(\textbf{Q}_e)$ and $\textbf{B}^v(\textbf{Q}_v)$, which act on the electron- and hole states. We now assume that the spin-orbit splittings are constants within both wave packets, i. e. they are given by characteristic but constant wave-vectors $\textbf{Q}_e$ and $\textbf{Q}_v$. 
  
 One usually distinguishes two important cases: In the case of direct excitons one considers the situation that $\textbf{Q}_e = \textbf{Q}_v = \textbf{Q}$, where $\textbf{Q}$ can be identified as the wave-vector of the critical point, at which the direct exciton is formed. In this case the single-particle wave-function packets from both bands forming the excitons are centered at the same point of the Brillouin zone. In the case of inter-valley excitons the critical points $\textbf{Q}_e$ and $\textbf{Q}_v$ denote the energy maximum of the valence band and at the energy minimum of the conduction band. They have to be specified but in this case $\textbf{Q}_e \neq \textbf{Q}_v$.

  Since conduction- and valence bands may both be modified via spin-orbit coupling, this gives rise to a spin-orbit contribution $H^{ex}_{so}$ to the exciton Hamiltonian, which takes the form:
  
   \begin{equation}  
  		H^{ex}_{so} =   a_{so}^{e}(\textbf{Q}_e)\sigma_e^z \otimes  1_v +  a_{so}^{v}(\textbf{Q}_v)  1_{e} \otimes \sigma_v^z   
  		\label{eqII13}	
  \end{equation}   
The sum of equ.~\eqref{eqII12} and equ.~\eqref{eqII13} determines the exciton spin-orbit Hamiltonian $H^{ex}_{dso}$ in the product space defined above (c.f. equ.~\eqref{eqII11}), and one obtains:

 \begin{equation}
 \begin{split} 
 H^{ex}_{dso} = H^{ex}_{d} + H^{ex}_{so} = a_{d} 1_{e} \otimes  1_v  + a_{so}^{e}(\textbf{Q}_e)\sigma_e^z \otimes  1_v +  a_{so}^{v}(\textbf{Q}_v)  1_{e} \otimes \sigma_v^z = \\
 = \left (
 \begin{array}{*{12}c}  
 a_{d} + a_{so}^{e}(\textbf{Q}_e) +a_{so}^{v}(\textbf{Q}_v) & 0 & 0 & 0 \\
 0 & a_{d} +  a_{so}^{e}(\textbf{Q}_e) - a_{so}^{v}(\textbf{Q}_v) & 0 & 0 \\
 0 & 0 &  a_{d} - a_{so}^{e}(\textbf{Q}_e) +a_{so}^{v}(\textbf{Q}_v)& 0 \\
 0 & 0 & 0 & a_{d} -  a_{so}^{e}(\textbf{Q}_e) -a_{so}^{v}(\textbf{Q}_v) \\
 \end{array}\right)
 \end{split} 
 \label{Hdso}
 \end{equation} 
 Thus, due to spin-orbit coupling the exciton states are no longer degenerate but obtain a fine structure \cite{JPEche_2016}. Since all contributions to the Hamiltonian operators that are  formulated in the sub-spaces of conduction- and valence-band electrons are given by diagonal matrices, $H^{ex}_{dso}$ is also diagonal.

\begin{figure}
	\begin{center}
		\includegraphics[width=15cm]{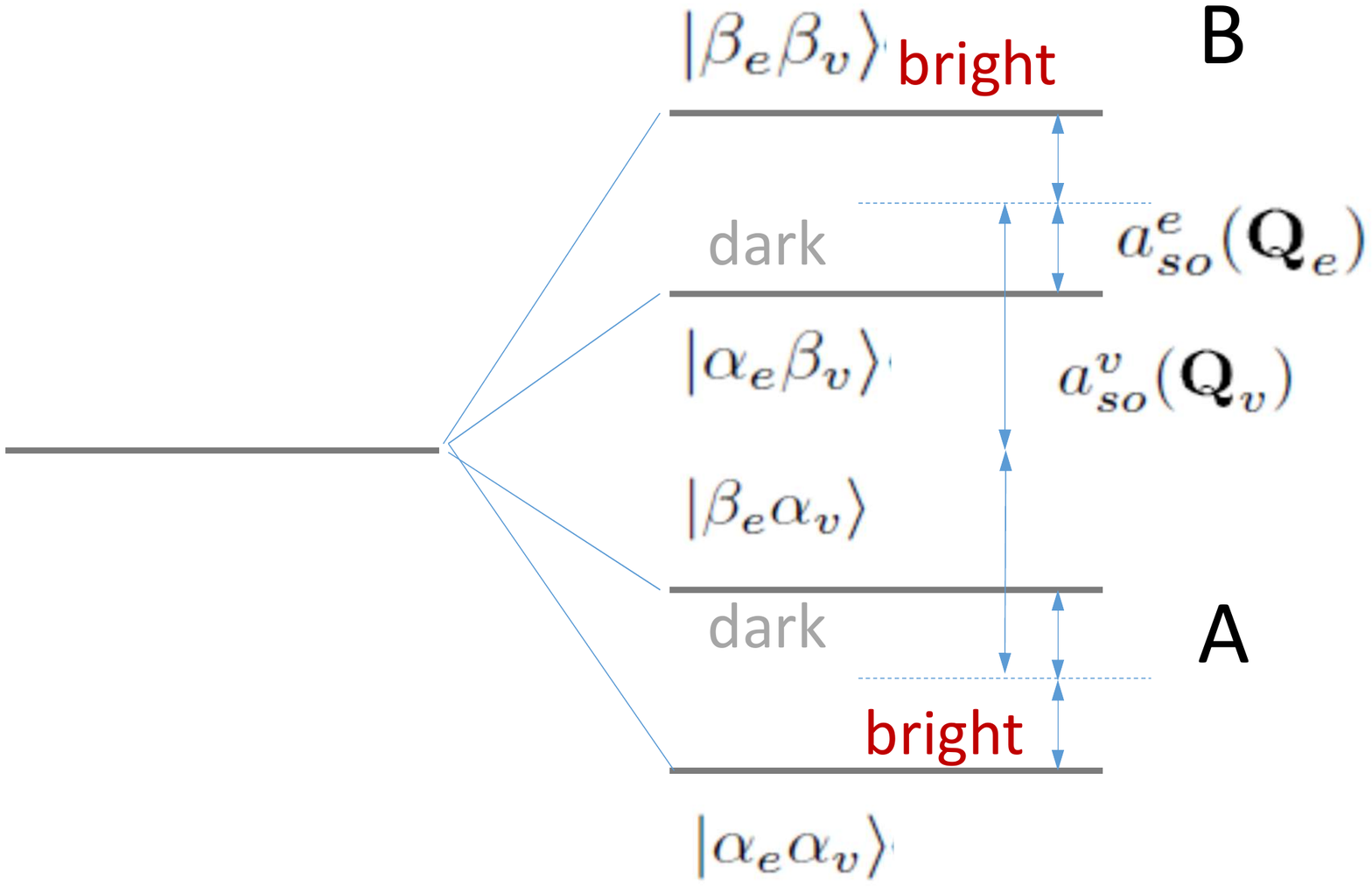}
	\end{center}
	\caption{Exciton structure resulting from the spin-orbit coupling. The labeling into  A and B excitons as well as into "bright" and "dark" exciton is given for the sake of illustration: Only the symmetry of full wave functions, including the orbital part and the spin allows to give the selection rules of the optical transitions and they can not be directly identified to the pseudo-spin we are dealing with in this work. Even, the absolute value and the sign of $a_{so}$ give  the right energetic order of the levels, that can not be determined from our model before further calculations. 
	}
	\label{figSOeffect}
\end{figure}

 The exciton  fine structure is further modified by the exchange interaction acting between conduction- and valence-band electron states. For sake of simplicity we neglect at the moment spin-orbit coupling and symmetry-breaking effects. The electron-hole exchange-interaction is due to the fact that electron and hole may be indistinguishable particles in the exciton wave function. This is the case the if spin- and wave-vector of the electron in the conduction band and that of the one, missing in the valence band are identical, i. e. if the quantum numbers of electron- and hole states (forming the exciton) are Kramer's conjugated to each other. 
   
  Three different terms (see Ref. \cite{Koster_1960})
 that have the symmetry properties of a Hamiltonian (i. e. that transform as ($\Gamma_1, \hat{K}^+$)) can now be constructed from the symmetry adapted basis matrices given in equ.~\eqref{eqII1}  and equ.~\eqref{eqII8}. The first, proportional to
  
  \[
  1_{e} \otimes  1_v  
 \]
 describes the direct Coulomb interaction between the conduction-band electron and the hole in the valence-band. It gives rise to the exciton binding energy introduced in~\eqref{eqII12}. 
 In addition, the Coulomb interaction gives rise to two exchange-interaction terms. They are given by: 
  \[  
 \sigma_e^z \otimes \sigma_v^z \text{   and   }  ( \sigma_e^{61} \otimes \sigma_v^{62}  + \sigma_e^{62} \otimes \sigma_v^{61} ) / \sqrt{2} 
 \]
The corresponding exciton-exchange interaction Hamiltonian $ H^{ex}_{ech} $ reads now explicitly:
 
 \begin{equation}
 H^{ex}_{ech} = a_{ech}^{zz} \sigma_e^z \otimes \sigma_v^z  +   a_{ech}^{66} ( \sigma_e^{61} \otimes \sigma_v^{62}  + \sigma_e^{62} \otimes \sigma_v^{61}) /\sqrt{2}
 \label{Hexech1}
 \end{equation}
where $a_{ech}^{zz}$ and $a_{ech}^{66}$ are constant numbers characterizing the exchange energy. Then  $ H^{ex}_{ech} $ has the matrix form in the basis of equ.~\eqref{eqII11}:
 
 \begin{eqnarray}   
 H^{ex}_{ech} & = &
   a_{ech}^{zz} \sigma_e^z \otimes \sigma_v^z  +   a_{ech}^{66} ( \sigma_e^{61} \otimes \sigma_v^{62}  + \sigma_e^{62} \otimes \sigma_v^{61}) /\sqrt{2}
   \nonumber
 \\
 & = & 
   \left (
 \begin{array}{*{12}c}  
  a_{ech}^{zz} & 0 & 0 & 0 \\
 0 & - a_{ech}^{zz} & -  a_{ech}^{66}/\sqrt{2}  & 0 \\
 0 &-   a_{ech}^{66}/\sqrt{2} & - a_{ech}^{zz} & 0 \\
 0 & 0 & 0 & a_{ech}^{zz} \\
 \end{array}\right)  
 \label{Hexech}
 \end{eqnarray} 
  Equ.~\eqref{eqII13} gives the only exchange interaction terms that can contribute to the Hamiltonian in this four-dimensional electron-hole subspace if symmetry-breaking interactions are neglected. No other terms have the required transformation properties ($\Gamma_1, \hat{K}^+$) for a Hamiltonian under the symmetry operations of the point group and under time reversal.

  As discussed in detail in Ref.~\onlinecite{Honer_invar_2018} or~\onlinecite{Hoener1980}, the exchange interaction has not a constant single value but may depend on external or internal perturbations. Such external perturbations may be electric or magnetic electric fields, strain or stress etc. An internal symmetry breaking perturbation can be the finite wave-vector $\bf Q$. These quantities have a well-defined symmetry. They are multiplied with the product of one basis Pauli matrices acting within the conduction-electron subspace and another one within the "missing-electron"-valence subspace. They thus modify the exciton exchange interaction. Besides the discussed transformation properties of the so constructed terms (as ($\Gamma_1, \hat{K}^+$)) in order to be present in a Hamiltonian) a necessary condition remains, however, that the involved electron states are characterized by the same set of quantum numbers. Wave-vector dependent exchange interaction can e. g. explain (see Ref.~\onlinecite{Honer_invar_2018}) the different effective masses of longitudinal and transverse excitons measured in CuCl~\cite{Mita1980}.
  
  Let us discuss the variation of the exchange interaction with wave-vector in the invariant expansion as an example. In this case the products of both Pauli matrices has to be multiplied with the center-of-mass wave-vector to a certain power and multiplied by a constant, indicating the strength of the interaction. Since in-plane wave-vectors as well as the Pauli matrices change their sign under time reversal, {\bf Q}-linear exchange interaction terms do not exist in our model for TMD mono-layers. The lowest order correction to the exchange interaction is thus given by a $Q^2 = (Q_x^2 + Q_y^2)$ -term multiplied by the exchange interaction given in eq.~\eqref{Hexech}. Both contributions are transforming separately according to ($\Gamma_1, \hat{K}^+$)) and thus also their product.

  At a level of development where other symmetry breaking interactions due to dispersion effects or external fields are not considered, the exciton Hamiltonian $H^{ex}$ is given by the sum of the exciton spin-orbit and exchange Hamiltonians (equ.~\eqref{Hdso} and equ.~\eqref{Hexech}):  
   
  \begin{equation}
     H^{ex} =  H^{ex}_{dso} +  H^{ex}_{ech}
  \end{equation}
     In order to study the emission- and absorption properties of a material it is interesting to consider its exciton fine-structure around its critical points. The solution of the Schrödinger equation for $H^{ex}$ in  equ. (II.17) gives this fine structure in a parameterized form, which will be discussed in the following.

  \section{  Fine Structure of the Exciton Ground State in  $MoSe_2$ Mono-Layers  }
  \setcounter{equation}{0}

    Transition-metal dichalcogenides mono-layers are direct gap materials. The energy gap is situated at the $K_{\pm}$-points that are characterized by their wave-vector $\textbf{K}_{\pm}$ and positioned at the edges of the two-dimensional hexagonal Brillouin zone \cite{KFMak_2010, DYQiu_2015}, the subscripts "${\pm}$" indicating valleys of different symmetry. The wave-vector group of the $K_{\pm}$ points is $C_{3h}$. As discussed above, our calculations considering spin-orbit coupling and exchange interaction can be (starting from the $\Gamma$-point) extended up to the $K_{\pm}$ points by introducing fictive magnetic fields $\textbf{B}^e(\textbf{Q}_e)$ and $\textbf{B}^v(\textbf{Q}_v)$ both $\parallel$ $\textbf{z}$. As indicated in  equ.~\eqref{Hdso} and~\eqref{Hexech} this calculation includes spin-orbit coupling and exchange-interaction terms of the conduction and valence band. 
    
    According to the spin structure of the considered bands our exciton basis consists of four states. Since in TMD mono layers the spin-orbit splitting of the valence-band states  at the $K_{\pm}$-points are usually much larger than the other symmetry-breaking interactions one observes two separated exciton series called "A" and "B" of packets of exciton states, which have an internal structure.  It is important to notice that terms originating from the spin-orbit coupling and those proportional to  $ a_{ech}^{zz}$ are diagonal terms in the basis of the states given in equ.~\eqref{eqII11}, i. e. they do not mix the wave functions. These terms account only for an energy re-normalization of the exciton states. Thus $\arrowvert\alpha_e\alpha_v\rangle$ and $\arrowvert\beta_e\beta_v\rangle$ are eigenstates of $H^{ex}$ but belong to different exciton series. The exchange-interaction term 
   \[
   a_{ech}^{66} ( \sigma_e^{61} \otimes \sigma_v^{62}  + \sigma_e^{62} \otimes \sigma_v^{61}) /\sqrt{2} 
   \]  
    mixes, on the contrary, the states $ \arrowvert\alpha_e\beta_v\rangle$ and $\arrowvert\beta_e\alpha_v\rangle$, which belong also to the different series. This exchange-interaction does not modify, however, the energies or wave functions of the states $\arrowvert\alpha_e\alpha_v\rangle$ and $\arrowvert\beta_e\beta_v\rangle$ given above.
    
       From the matrix equations~\eqref{Hdso} and~\eqref{Hexech} one determines the energies $E_{A1}^{ex}$ and $E_{A2}^{ex}$ ($E_{B1}^{ex}$  and $E_{B2}^{ex}$) of the exciton states of the A (B) series to:

   \begin{equation}
   \begin{split}
   E_{B1}^{ex} =  a_{d} +  a_{ech}^{zz} + a_{so}^{e}(\textbf{Q}_e) + a_{so}^{v}(\textbf{Q}_v)  \\
   E_{A1}^{ex} =  a_{d}  +  a_{ech}^{zz} - a_{so}^{e}(\textbf{Q}_e) - a_{so}^{v}(\textbf{Q}_v) \\  
   \end{split}
   \label{eqIII1}
   \end{equation}    
    for the states $\arrowvert\alpha_e\alpha_v\rangle$ and $\arrowvert\beta_e\beta_v\rangle$, respectively, and for the mixed states the energies:
   
   \begin{equation}
   \begin{split}
   E_{B2}^{ex} =  a_{d} -  a_{ech}^{zz} + \sqrt{ (- a_{so}^{e}(\textbf{Q}_e) + a_{so}^{v}(\textbf{Q}_v))^2 +  (a_{ech}^{66})^2 / 2   } \\
   E_{A2}^{ex} =   a_{d} -  a_{ech}^{zz} - \sqrt{(- a_{so}^{e}(\textbf{Q}_e) + a_{so}^{v}(\textbf{Q}_v))^2 +  (a_{ech}^{66})^2 / 2   }  \\  
   \end{split}
   \label{eqIII2}
   \end{equation}
   
   The energy splitting

   \begin{equation}
   \Delta_{AB} = E_{A1}^{ex} -  E_{B1}^{ex} = 2( a_{so}^{e}(\textbf{Q}_e) +a_{so}^{v}(\textbf{Q}_v)  
   \label{eqIII3}
   \end{equation} 
   corresponds to the main contribution to the energy separation between the A and B exciton series.  
   
     Equs.~\eqref{eqII1} and~\eqref{eqII2} show that the mixed states are not degenerate with the other exciton states but have slightly different energies. This is mainly caused by the spin-orbit splitting of the conduction band and by the different contributions of the exchange interactions. This point will be discussed in detail in chapter III.A and III.B for direct and inter-valley excitons.
   
   If a magnetic field $B^z$ is applied perpendicularly to the mono-layer, the exciton fine structure is not further changed.  In the case of small magnetic fields (linear Zeeman effect), only their energies are slightly modified. The corresponding eigenvalues are obtained by replacing in equ. (III.1) and (III.2): 
   
   \[
   \begin{split}
   a_{so}^{e}(\textbf{Q}_e)  \text{ by }  (a_{so}^{e}(\textbf{Q}_e) + g^e \mu_B B^z ) \\
   \text{and } \\
   a_{so}^{v}(\textbf{Q}_v)  \text{ by }  (a_{so}^{v}(\textbf{Q}_v) + g^v \mu_B B^z ) \\  
   \end{split}
   \]  
   Here $g^e$ and $g^v$ denote the Landé factors of the conduction- and valence-band electrons, respectively. For higher fields all exciton energies $a_{d}$ in ~\eqref{eqIII1} and~\eqref{eqIII2} are shifted proportional to  $(B^z)^2$ due to a quadratic Zeeman effect. 
   Applying an in-plane magnetic field $\textbf{B} = [B^x, B^y, 0]$ leads, however, to a complex mixing of all exciton states, which is not further discussed here.

  \subsection{  Direct or Intra-Valley Excitons }

   \setcounter{equation}{0}

   In optical absorption processes an electromagnetic radiation field (or photon) excites through its dipole moment an electron from an occupied valence-band state to an unoccupied conduction-band state. The transition takes place with conservation of energy and momentum, i. e. the energy of the photon $\hbar\omega$ is transferred to the excited electron and the photon wave-vector $\textbf{q}$ added to that of the valence band electron $\textbf{Q}_v$. One thus obtains:

   \begin{equation}
   \begin{split}
   E^{e} =   E^{v} + \hbar \omega \\
   \text{and } \\
    \textbf{Q}_e = \textbf{Q}_v + \textbf{q}\\  
   \end{split}
   \end{equation}
   Normally the photon wave-vector is negligibly small compared to that of the valence-band electrons from which the exciton states are constructed and $\textbf{Q}_e = \textbf{Q}_v$ holds, leading to the expression "direct excitons". Only such direct excitons are accessible in optical processes if no other quasi-particles (as phonons or crystal imperfections) are involved. The  probability or quantum yield of the transition is determined by its dipole-matrix element.
   
    In addition to the wave-vector conservation only the orbital part of the electron-wave function is modified throughout the transition, but the electron spin is conserved in the optical excitation process. Such transitions are possible for the  direct exciton states $\arrowvert\alpha_e\alpha_v\rangle$ and $\arrowvert\beta_e\beta_v\rangle$ of our four-level model. These excitons are thus optical active and called "bright" excitons (subscript "b" in the following). The states $ \arrowvert\alpha_e\beta_v\rangle$ and $\arrowvert\beta_e\alpha_v\rangle$ and their mixture due to exchange interaction are called "dark" (subscript "d") states. In their excitation (although direct) the electron spin has to change, which is not possible when the transition is induced by an electromagnetic radiation field alone. Transitions to dark states involve a spin-flip of the electron, which needs the presence of an additional perturbation.  
   
  In general exciton states are constructed by introducing hole states instead of the valence-band states. These states are Kramer's conjugated states of each other. Then, in exciton notation, the bright state $\arrowvert\alpha_e\alpha_v \rangle$ transforms into  $\arrowvert\alpha_e\beta_h\rangle$ in which electron- and hole spins are anti-parallel. In analogy with atomic physics this electron-hole pair is called to be in a "spin-singlet" state. In addition, the hole wave-vector $\textbf{Q}_h$ is related to $\textbf{Q}_v$ by Kramer's conjugation through:

    \begin{equation}
    \textbf{Q}_h = - \textbf{Q}_v    
   \end{equation}
   This leads together with equ. (III.A1) to the fact that 
    \begin{equation}
   \textbf{Q}_e + \textbf{Q}_h = 0  
   \label{eqIIIA3}  
   \end{equation}
 for optical active excitons, i. e. they are situated at the $\Gamma$-point of the exciton momentum space.

  Our calculations includes spin-orbit and exchange-interaction terms, where in the present case conduction- and valence-electron states forming the exciton have the same wave-vector $\textbf{Q}$, i. e. they are situated within the same energy valley \cite{TYu_2014}. The critical points are the points $K_{+}$ and $K_{-}$, respectively, characterized by their wave-vectors $\textbf{K}_{+}$ and $\textbf{K}_{-}$. Introducing the notation $\textbf{K}_{+}$ = $\textbf{K}$ and using the symmetry relation $\textbf{K}_{-}$ = - $\textbf{K}_{+}$ =  - $\textbf{K}$ we obtain from equ. (III.1) the energies:

\begin{equation}
\begin{split}
E_{bB}^{ex} =  a_{d} +  a_{ech}^{zz} + a_{so}^{e}(\textbf{K}_) + a_{so}^{v}(\textbf{K})  \\
E_{bA}^{ex} =  a_{d}  +  a_{ech}^{zz} - a_{so}^{e}(\textbf{K}) - a_{so}^{v}(\textbf{K}) \\  
\end{split}
   \label{eqIIIA4}
\end{equation}
for the bright exciton states at the $K_{+}$ point of the electron Brillouin zone. As in equ.~\eqref{eqIII3}

 \[
\Delta_{AB} = E_{bA}^{ex} -  E_{bB}^{ex} = 2( a_{so}^{e}(\textbf{K}) +a_{so}^{v}(\textbf{K})  
\]
is the energy splitting between the A and B exciton series. 

Concerning the spin-triplet exciton states (the dark states, i. e. states where the spins of electron and hole are parallel) their energies $E_{dA}^{ex}$ and $E_{dB}^{ex}$ are similar to the bright states obtained from equ.~\eqref{eqIII2}. Since in most materials the spin-orbit splitting of the valence-band states is large compared to all other energy re-normalization effects, bright- and dark-exciton states are lying pairwise closely together in the A- and B-exciton bands. We now write equ.~\eqref{eqIII2} for the dark states in the form:
\begin{equation}
\begin{split}
  E_{dB}^{ex} =  a_{d} -  a_{ech}^{zz} + a_{so}^{v}(\textbf{K}) \sqrt{ (1 -  a_{so}^{e}(\textbf{K})/ a_{so}^{v}(\textbf{K}))^2 +  (a_{ech}^{66}/ a_{so}^{v}(\textbf{K}))^2 / 2   }
   \\
  E_{dA}^{ex} =   a_{d} -  a_{ech}^{zz} - a_{so}^{v}(\textbf{K}) \sqrt{ (1 -  a_{so}^{e}(\textbf{K})/ a_{so}^{v}(\textbf{K}))^2 +  (a_{ech}^{66}/ a_{so}^{v}(\textbf{K}))^2 / 2   } \\
\end{split}
   \label{eqIIIA5}
\end{equation}
  and develop the square-root expressions in equs.~\eqref{eqIIIA5} up to terms of first order in $1/ |a_{so}^{v}(\textbf{K})|$. One then obtains:

   \begin{equation}
   \begin{split}
   E_{dB}^{ex} =  a_{d} -  a_{ech}^{zz} + a_{so}^{v}(\textbf{K}) - 
    a_{so}^{e}(\textbf{K}) + (a_{so}^{e}(\textbf{K})) ^2 / (2 |a_{so}^{v}(\textbf{K})|) +  (a_{ech}^{66}) ^2 / (4 |a_{so}^{v}(\textbf{K})|)    \\
   E_{dA}^{ex} =   a_{d} -  a_{ech}^{zz} - a_{so}^{v}(\textbf{K}) + 
   a_{so}^{e}(\textbf{K}) - (a_{so}^{e}(\textbf{K})) ^2 / (2 |a_{so}^{v}(\textbf{K})|) -  (a_{ech}^{66}) ^2 / (4 |a_{so}^{v}(\textbf{K})|)   \\  
   \end{split}
   \label{eqIIIA6}
   \end{equation}
     Comparing the energies given in equ.~\eqref{eqIIIA6} to that of equ.~\eqref{eqIIIA4}, bright and dark exciton states of the A-series are separated in energy by:
  
    \begin{equation}
     \Delta_{Abd} = E_{bA}^{ex} -  E_{dA}^{ex} = 2(a_{ech}^{zz} + a_{so}^{e}(\textbf{K})) - (a_{so}^{e}(\textbf{K})) ^2 / (2 |a_{so}^{v}(\textbf{K})|) -  (a_{ech}^{66}) ^2 / (4 |a_{so}^{v}(\textbf{K})|)  
     \label{eqIIIA7}
  \end{equation} 
   Within the B-series the energy separation is given by:
  
   \begin{equation}
   \Delta_{Bbd} = E_{bB}^{ex} -  E_{dB}^{ex} =  2(a_{ech}^{zz} - a_{so}^{e}(\textbf{K}))  + (a_{so}^{e}(\textbf{K})) ^2 / (2 |a_{so}^{v}(\textbf{K})|) +  (a_{ech}^{66}) ^2 / (4 |a_{so}^{v}(\textbf{K})|)     
        \label{eqIIIA8}
  \end{equation}
Equ.~\eqref{eqIIIA7} and equ.~\eqref{eqIIIA8} show that the fine structure of the A- and B-exciton series are different due to the spin-orbit interaction of the conduction-band electrons and the nondiagonal exchange interaction term $a_{ech}^{66}$. 

\begin{figure}
	\begin{center}
		\includegraphics[width=15cm]{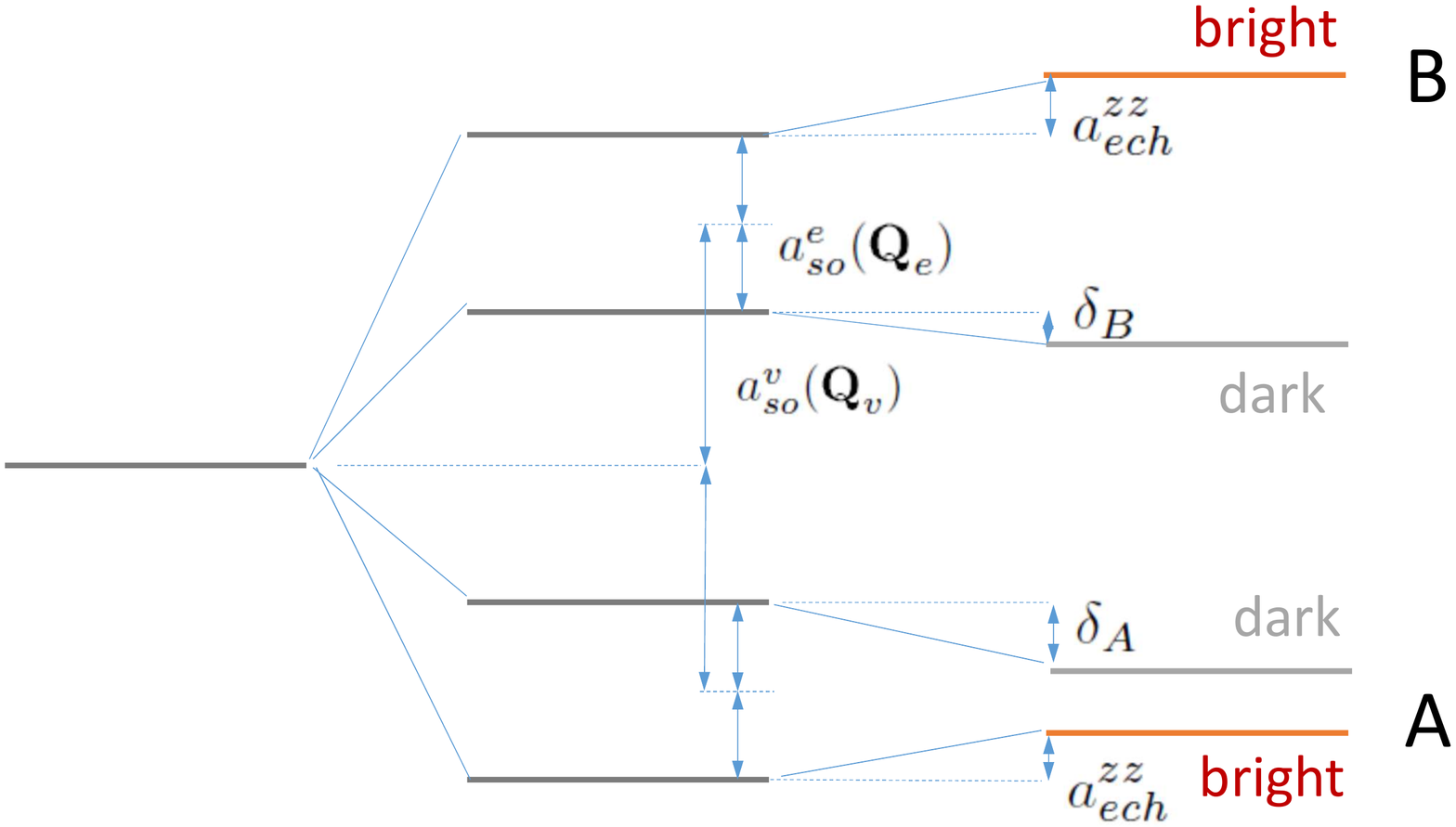}
	\end{center}
	\caption{Fine structure of the intra-valley excitons as resulting from the spin-orbit coupling and from the electron-hole exchange interaction. The $\delta$ parameters are defined as
		$
		\delta_{B} 
		a_{ech}^{zz} -(a_{so}^{e}(\textbf{K})) ^2 / (2 |a_{so}^{v}(\textbf{K})|) -  (a_{ech}^{66}) ^2 / (4 |a_{so}^{v}(\textbf{K})|)  
		$
		and
		$
		\delta_{A} 
		a_{ech}^{zz} +(a_{so}^{e}(\textbf{K})) ^2 / (2 |a_{so}^{v}(\textbf{K})|) -  (a_{ech}^{66}) ^2 / (4 |a_{so}^{v}(\textbf{K})|)  
		$
		The remark about the labeling, the selection rules and the spectral positions of the levels given previously is still valid here. 
	}
	\label{figExcIntravalley}
\end{figure}

The same energies as given in eqs.~\eqref{eqIIIA4} and~\eqref{eqIIIA5} are obtained for the dark- and bright excitons if the electron in the conduction band and the missing electron in the valence band are both situated at the $K_{-}$ points of the electron Brillouin zone. In this case, because of the symmetry relation discussed above, $a_{so}^{e}(-\textbf{K})$ = - $a_{so}^{e}(\textbf{K})$ and $a_{so}^{v}(-\textbf{K})$ = - $a_{so}^{v}(\textbf{K})$ hold. The energies obtained now belong, however, to the bright- or dark-exciton states interchanged with respect to the  $K_{+}$ critical points.

  Besides the exciton-fine structure, equ.~\eqref{eqIIIA6} also shows that the two contributions to the exchange interaction  $a_{ech}^{zz}$ and $a_{ech}^{66}$  have different importance: While the energy re-normalization proportional to $a_{ech}^{zz}$ is independent of the spin-orbit interaction, the one proportional to $a_{ech}^{66}$ becomes negligible for very important spin-orbit splittings of the valence band states. In addition, equ.~\eqref{IIIA5} shows that the term proportional to $a_{ech}^{66}$ may become important (leading to an appreciable mixing of the dark states) if the spin-orbit splittings of conduction-band electrons and that of the holes in the valence band diminish or if they compensate each other. Then the energies of the dark states become independent of the spin-orbit interactions and may be very different from that of the bright states.

   \subsection{Inter-Valley  Excitons }

\setcounter{equation}{0}

  The construction of excitons from occupied conduction- and vacant valence-band electron states discussed in chapter II is quite general and uses only the symmetry properties of the crystal and the spin structure of the degenerate states. Therefore it can also be applied to the construction of inter-valley exciton states, i. e. in situations where equ.~\eqref{IIIA3} does not apply. In TMD mono-layers the direct gap is situated at the $K_{+}$ and $K_{-}$ points of the electron Brillouin zone where their energy is degenerated. In addition, opposite $K_{+}$ and $K_{-}$ points are separated by a reciprocal lattice vector and their corresponding wave-vectors are therefore related to each other by time-reversal symmetry. In this situation, the Coulomb interaction introduced above leads also to exchange interaction terms between electrons and holes from different $K_{\pm}$ valleys. This interaction is called "inter-valley exchange interaction". The valleys are, however, not completely equivalent to each other since the energetic order of electron- and hole states is reversed in the two valleys and the states may have different symmetries at high symmetry points.
  
  If exciton states have different symmetry, they may obey to different selection rules for optical transitions. This is the case for the $K_{+}$ and $K_{-}$ critical points in TDM mono-layers. The bright states are optically active when exciting with polarized light. In this case the different valleys can be excited selectively by employing right- or left-hand circularly polarized light. Outside the $K_{\pm}$ points the inter-valley exchange interaction results then into a mixing of exciton states from different valleys. Thus the symmetry of the exciton states is modified and the optical selection rules of the different valleys are relaxed, leading to an inter-valley transfer of electron-hole excitation \cite{CRobe_2017}.
 
 The influence of inter-valley exchange on bright- and dark-exciton states may be treated e. g. in the framework of $\textbf{k}$ $\cdot$ $\textbf{p}$ perturbation theory (see Ref. \cite{MMGlazov_2014}). In addition as we will discuss here in the following, the exciton fine structure can be modified due to the inter-valley exchange interaction and the electron spin-orbit coupling. 
  
 Let us consider first the case that an inter-valley exciton is formed from a conduction-band electron in the $K_{-}$-valley and a missing valence-band electron in the $K_{+}$-valley. Under these conditions an exciton is formed at the  $K_{-}$-point of the exciton Brillouin zone. As discussed above, these excitons cannot be optically excited (i. e. they cannot be classified as "dark" or "bright"), but their spin-structure is still identified as spin-singlet (subscript "s") or spin-triplet (subscript "t").

 Since the critical points of the $K_{-}$ and  $K_{+}$ valleys are characterized by the opposite wave vectors $\textbf{K}_-$ and $\textbf{K}_+$, we obtain from the symmetry condition of the spin-orbit interaction for the conduction band electrons at the critical point  $K_{-}$:
 
 \begin{equation}
 a_{so}^{e}(-\textbf{K}) = - a_{so}^{e}(\textbf{K}) 
 \label{eqIIIB1}
 \end{equation}
  where we have used the notation $\textbf{K}_{-}$ =   -$\textbf{K}$ introduced above. The value given in  equ.~\eqref{eqIIIB1} has now to be used in equ.~\eqref{eqIII1} and~\eqref{eqIII2}. The valence-band electron spin-orbit interaction remains unchanged since it is supposed to be in the $K_{+}$-valley as in the situation discussed in chapter III.A. We thus obtain similar to equ. (III.A4) for the inter-valley singlet excitons:

 \begin{equation}
 \begin{split}
 E_{Bs}^{ex} =  a_{d} +  a_{ech}^{zz} - a_{so}^{e}(\textbf{K}) + a_{so}^{v}(\textbf{K})  \\
 E_{As}^{ex} =  a_{d}  +  a_{ech}^{zz} + a_{so}^{e}(\textbf{K}) - a_{so}^{v}(\textbf{K}) \\  
 \end{split}
 \label{eqIIIB2}
 \end{equation}
 where as in equ.~\eqref{eqIII3}
 \[
 \Delta_{ABs} = E_{As}^{ex} -  E_{Bs}^{ex} = 2(- a_{so}^{e}(\textbf{K}) + a_{so}^{v}(\textbf{K}) ) 
 \]
 is the energy splitting between the inter-valley singlet A and B exciton states. For the inter-valley triplet excitons we obtain analog to equ.~\eqref{eqIIIA5}
 
\begin{eqnarray}
 E_{Bt}^{ex}
  & = & 
  a_{d} -  a_{ech}^{zz} + a_{so}^{v}(\textbf{K}) \sqrt{ (1 +  a_{so}^{e}(\textbf{K})/a_{so}^{v}(\textbf{K}))^2 +  (a_{ech}^{66}/ a_{so}^{v}(\textbf{K}))^2 / 2   } 
 \nonumber
 \\
 E_{At}^{ex} 
 & = &  a_{d} -  a_{ech}^{zz} - a_{so}^{v}(\textbf{K}) \sqrt{ (1 +  a_{so}^{e}(\textbf{K})/a_{so}^{v}(\textbf{K}))^2 +  (a_{ech}^{66}/ a_{so}^{v}(\textbf{K}))^2 / 2   }  
 \label{eqIIIB3}
\end{eqnarray}
 After developing the square-root expressions in equs.~\eqref{eqIIIB3} one obtains:
\begin{eqnarray}
E_{Bt}^{ex}
& = & 
a_{d} -  a_{ech}^{zz} + a_{so}^{v}(\textbf{K}) + 
a_{so}^{e}(\textbf{K}) + (a_{so}^{e}(\textbf{K})) ^2 / (2 |a_{so}^{v}(\textbf{K})|) +  (a_{ech}^{66}) ^2 / (4 |a_{so}^{v}(\textbf{K})|) 
\nonumber
\\
E_{At}^{ex} 
& = &
a_{d} -  a_{ech}^{zz} - a_{so}^{v}(\textbf{K}) - 
a_{so}^{e}(\textbf{K}) - (a_{so}^{e}(\textbf{K})) ^2 / (2 |a_{so}^{v}(\textbf{K})|) -  (a_{ech}^{66}) ^2 / (4 |a_{so}^{v}(\textbf{K})|)   
\label{eqIIIB4}
\end{eqnarray}
 
\begin{figure}
	\begin{center}
		\includegraphics[width=15cm]{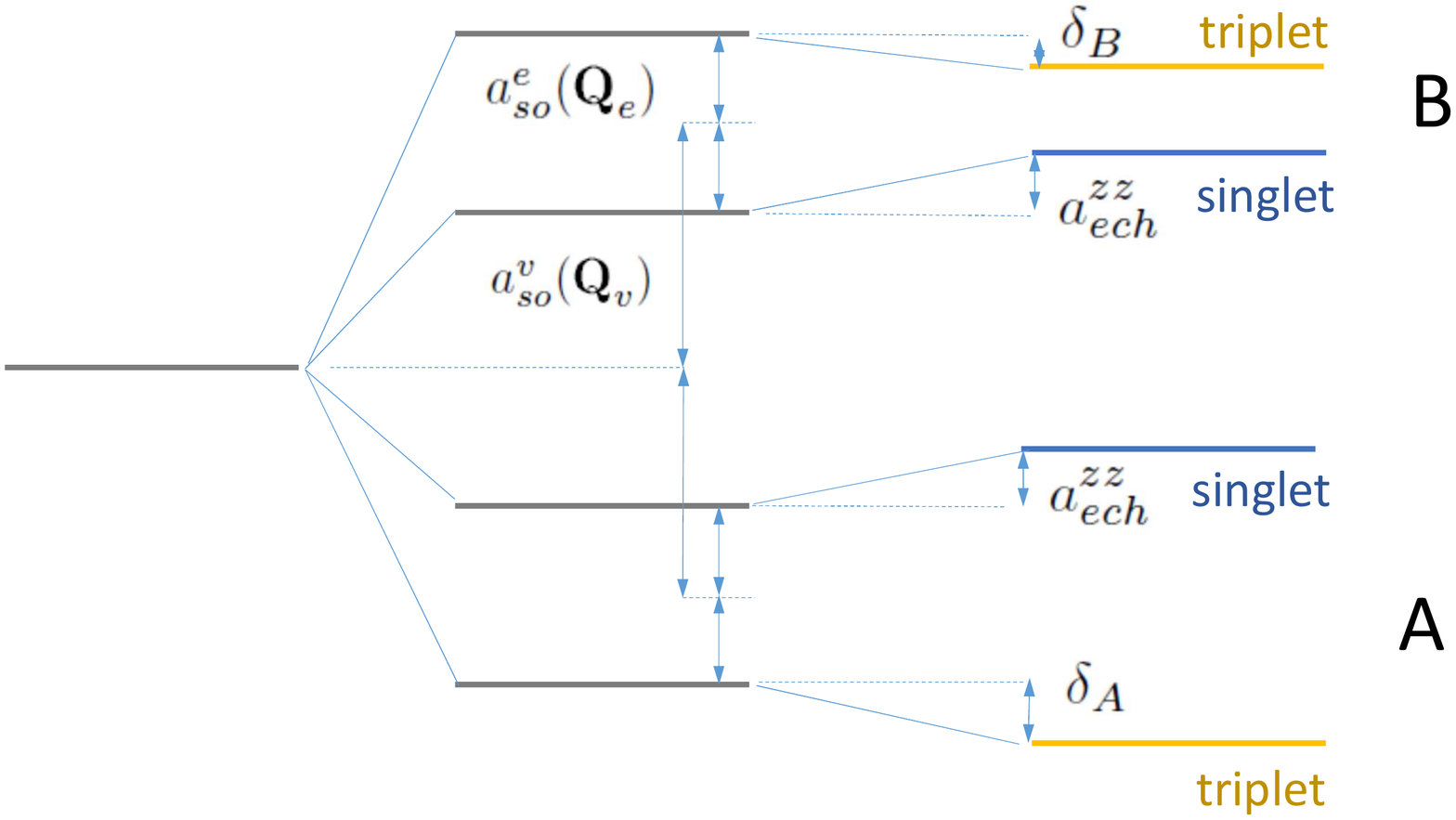}
	\end{center}
	\caption{Fine structure of the inter-valley excitons as resulting from the spin-orbit coupling and from the electron-hole exchange interaction. The $\delta$ parameters are defined as for Fig.~\ref{figExcIntravalley}
	}
\end{figure}

The same energies as given in eqs. (III.B2) and (III.B4) are obtained for the singlet- and triplet excitons if the electron in the conduction band and the missing electron in the valence band are situated in the $K_{+}$ and $K_{-}$-valleys of the electron Brillouin zone, respectively. The two singlet- or triplet-exciton states for this valley configuration are, however, interchanged when compared to the former situation.

   \section{ Fine Structure of the A - Exciton Ground State }

\setcounter{equation}{0}

To illustrate our results, we can have a closer look onto the fine structure of the A - exciton ground state.

 In the TMD mono-layers discussed above the electrons and holes are characterized by two spin states. Each of both quasi-particles can be situated close to two different critical points at which a direct energy gap opens in the electronic band structure. When the exciton states are formed through electron-hole pairs which can be characterized by the spin- and valley index of each of the constituents. Thus 16 exciton states can be formed. If the spin-orbit splitting of the valence band is big compared to all other interactions, two packets of eight different states are energetically close together, forming the ground state of the "A" and "B" exciton series. Within these packets the exciton states show a fine structure. These states are pairwise degenerate and are labeled as "bright (b)", "dark (d)", "singlet (s)", and "triplet (t)" states.
 
 It is interesting to consider e. g. the fine structure of the A-exciton states (having the lower energy) in detail, neglecting (compared to our former discussion) all terms of the order of $1/|a_{so}^{v}(\textbf{K})|$. With respect to their common energy value: 
 \[
 E_0 = a_{d}  + a_{so}^{v}(\textbf{K}) 
 \]
 the A-exciton states are subject to the energy shifts: 
  
 \begin{equation}
 \begin{split}
  \text{For the direct excitons } \\
 E_{bA}^{} = a_{ech}^{zz} + a_{so}^{e}(\textbf{K})  \text{ and }  E_{dA}^{} =  -  a_{ech}^{zz} -  a_{so}^{e}(\textbf{K}) \\
  \text{and for the inter-valley excitons } \\
  E_{As}^{} =  a_{ech}^{zz} - a_{so}^{e}(\textbf{K}) \text{ and } E_{At}^{} =  -  a_{ech}^{zz} +  a_{so}^{e}(\textbf{K}), \\
 \end{split}
 \end{equation}
  where each term is doubly degenerated, depending on the spin structure of the electron- and hole states in the different valleys. We see that a complex energy-fine structure of the different exciton states is already obtained if the diagonal exchange interaction term $a_{ech}^{zz}$ and the electron spin-orbit coupling $a_{so}^{e}(\textbf{K})$ are considered. It depends on the spin- and valley structure of the constituents. When including in addition the non-diagonal exchange interaction term $a_{ech}^{66}$, the energy shifts given above are modified and the exciton wave functions of the dark- triplet A- and B-exciton states are mixed.
 
    \section{Conclusion}
    
 The invariant expansion of the Hamiltonian is a fruitful method that was already proved to be very efficient for the study of the usual Zinc-blende- and Wurtzite-type semiconductor crystals~\cite{Honer_invar_2018}. We show here that it can give also valuable information about the fine structure of electronic excitations in layered TMD materials.    
  The model does not allow to give the value of the parameters describing the spin-orbit and exchange interaction nor the exact position of the exciton levels that depends on the sign and magnitude of those parameters. Nevertheless, starting from very general considerations about the symmetries of the crystals, it gives the exact fine structure of excitons that quantitative calculations must find. Moreover it shows that  an interaction such as the electron-hole exchange can not be considered as only shifting the level position, but that it gives rise also to non-diagonal terms in the Hamiltonian that couple the levels and allows for transfers between them.     
 
This invariant expansion method is very versatile and could be extended to other problem interesting the TMDs: mixed excitons coupling the conduction band at the $\Gamma$-point and the valence band at the K point, fine structure of biexcitons made of various types of excitons, effect of an external perturbation due for example to an applied magnetic field, etc.
 
\newpage

	\section*{ References}
	\bibliography{Pdichal-A-B-ex}

\end{document}